\newif\ifisdouble

\isdoubletrue 

\ifisdouble
	\documentclass{article}
	\usepackage{spconf}
\else
	\documentclass[11pt,draftcls,onecolumn]{IEEEtran}
\fi

\usepackage[utf8]{inputenc}

\usepackage[english]{babel}
\usepackage{blindtext}

\usepackage{subfigure}

\usepackage{amsmath, amsthm, url}
\usepackage{amssymb}
\usepackage{latexsym}

\usepackage{balance}

\usepackage[defblank]{paralist}

\usepackage[ruled]{algorithm2e}

\usepackage{dashrule}

\SetAlFnt{\small}
\SetAlCapFnt{\small}
\SetAlCapNameFnt{\small}
\SetAlCapHSkip{0pt}
\IncMargin{-\parindent}

\usepackage{ifpdf}

\ifpdf
\usepackage[pdftex]{graphicx}
\else
\usepackage{graphicx}
\fi

\ninept

\title{HYBRID CODING OF VISUAL CONTENT AND LOCAL IMAGE FEATURES}

\name{Luca Baroffio, Matteo Cesana, Alessandro Redondi, Marco Tagliasacchi, Stefano Tubaro\thanks{The project GreenEyes acknowledges the financial support of the Future and Emerging Technologies (FET) programme within the Seventh Framework Programme for Research of the European Commission, under FET-Open grant number:296676.}}
\address{Dipartimento di Elettronica, Informazione e Bioingegneria, Politecnico di Milano}

\begin{document}

\ifpdf
\DeclareGraphicsExtensions{.pdf, .jpg, .tif}
\else
\DeclareGraphicsExtensions{.eps, .jpg}
\fi

\newcommand{\x}{\mathbf{x}}
\newcommand{\xt}{\tilde{\x}}
\newcommand{\xh}{\hat{\x}}
\newcommand{\xb}{\bar{\x}}
\newcommand{\X}{\mathbf{X}}
\newcommand{\Xh}{\hat{\X}}
\newcommand{\Xt}{\tilde{\X}}

\newcommand{\Descr}{\mathcal{D}}
\newcommand{\Descrt}{\tilde{\Descr}}

\newcommand{\descr}{d}
\newcommand{\descrq}{\tilde{\descr}}
\newcommand{\descrt}{\tilde{\descr}}

\newcommand{\dv}{\mathbf{\descr}}
\newcommand{\dq}{\tilde{\dv}}
\newcommand{\dvt}{\mathbf{\tilde{\dv}}}

\newcommand{\cv}{\mathbf{c}}
\newcommand{\cq}{\tilde{\cv}}

\newcommand{\Dim}{D}

\newcommand{\scale}{\sigma}
\newcommand{\scalet}{\tilde{\scale}}

\newcommand{\orient}{\theta}
\newcommand{\coord}{\mathbf{p}}
\newcommand{\coordq}{\tilde{\coord}}

\newcommand{\Rate}{R}
\newcommand{\Dist}{D}

\newcommand{\Image}{\mathcal{I}}
\newcommand{\Keypoints}{\mathcal{K}}
\newcommand{\ImageDec}{\tilde{\mathcal{I}}}

\newcommand{\T}{\mathbf{T}}

\newcommand{\ATC}{\texttt{ATC}}
\newcommand{\HATC}{\texttt{HATC}}
\newcommand{\HATCfull}{\textit{``Hybrid-Analyze-Then-Compress''}}
\newcommand{\ATCfull}{\textit{``Analyze-Then-Compress''}}
\newcommand{\CTA}{\texttt{CTA}}
\newcommand{\CTAfull}{\textit{``Compress-Then-Analyze''}}

\def\codeif{\mbox{\textbf{if\ }}}
\def\codefi{\mbox{\textbf{fi\ }}}
\def\codethen{\mbox{\textbf{then\ }}}
\def\codeforeach{\mbox{\textbf{for each\ }}}
\def\codefor{\mbox{\textbf{for\ }}}
\def\codedo{\mbox{\textbf{do\ }}}
\def\codewhile{\mbox{\textbf{while\ }}}
\def\codetrue{\mbox{\textbf{true\ }}}
\def\codefalse{\mbox{\textbf{false\ }}}
\def\codereturn{\mbox{\textbf{return\ }}}
\def\codebreak{\mbox{\textbf{break\ }}}
\def\codebool{\mbox{\textbf{bool\ }}}
\def\codeelse{\mbox{\textbf{else\ }}}
\def\codeswitch{\mbox{\textbf{switch\ }}}
\def\codecase{\mbox{\textbf{case\ }}}
\def\codeand{\mbox{\textbf{and\ }}}
\def\codeend{\mbox{\textbf{end\ }}}

\newcommand{\ind}{\quad}
\newcommand{\tallind}{\quad\quad}

\maketitle

\begin{abstract}
Distributed visual analysis applications, such as mobile visual search or Visual Sensor Networks (VSNs) require the transmission of visual content on a bandwidth-limited network, from a peripheral node to a processing unit. Traditionally, a ``Compress-Then-Analyze" approach has been pursued, in which sensing nodes acquire and encode the pixel-level representation of the visual content, that is subsequently transmitted to a sink node in order to be processed. This approach might not represent the most effective solution, since several analysis applications leverage a compact representation of the content, thus resulting in an inefficient usage of network resources. Furthermore, coding artifacts might significantly impact the accuracy of the visual task at hand. To tackle such limitations, an orthogonal approach named ``Analyze-Then-Compress" has been proposed~\cite{CTA_ATC}. According to such a paradigm, sensing nodes are responsible for the extraction of visual features, that are encoded and transmitted to a sink node for further processing. In spite of improved task efficiency, such paradigm implies the central processing node not being able to reconstruct a pixel-level representation of the visual content. In this paper we propose an effective compromise between the two paradigms, namely ``Hybrid-Analyze-Then-Compress" (HATC) that aims at jointly encoding visual content and local image features. Furthermore, we show how a target tradeoff between image quality and task accuracy might be achieved by accurately allocating the bitrate to either visual content or local features.  
\end{abstract}


\section{Introduction}
\label{sec:intro}
In the last few years, local features have been effectively exploited in a number of visual analysis tasks such as augmented reality, object recognition, content based retrieval, image registration, etc. They provide a robust yet concise representation of an 
image patch that is invariant to local and global transformation such as illumination and viewpoint changes. The traditional pipeline for the extraction of local image feature consists of two main stages: i) a keypoint detector, that aims at identifying salient points within an image and ii) a keypoint descriptor that captures the local information of the image patch surrounding each keypoint. Traditional algorithms for keypoint description, such as SIFT~\cite{DBLP:journals/ijcv/Lowe04} and SURF~\cite{DBLP:conf/eccv/BayTG06}, assign to each salient point a description by means of a set of real-valued elements, capturing local information based on intensity gradient. More recently, a novel class of algorithms, namely binary descriptors, has emerged as an effective, yet computationally efficient, alternative to SIFT and SURF. Such features usually rely on smoothed pixel intensities and not on local intensity gradients, vastly improving the computational efficiency. The BRIEF~\cite{DBLP:conf/eccv/CalonderLSF10} descriptor consists of a set of binary values, each obtained by comparing the smoothed intensity of two pixels, randomly sampled around a keypoint. BRISK~\cite{DBLP:conf/iccv/LeuteneggerCS11}, ORB~\cite{Rublee_ORB} and FREAK~\cite{DBLP:conf/cvpr/AlahiOV12} refine the process, introducing ad-hoc designed spatial patterns of pixels to be compared and achieving rotation-invariance. More recently, BAMBOO~\cite{BAMBOO_MMSP}\cite{BAMBOO_ICASSP} exploits a pairwise boosting algorithm to build a discriminative pattern of pairwise pixel intensity comparisons.

Local features represent a key component of many distributed visual analysis applications such as Mobile Visual Search, augmented reality, and Visual Sensor Networks applications. Traditionally, such tasks have been tackled according to a \emph{Compress-Then-Analyze} ($\CTA$) approach, in which sensing nodes acquire the content, encode it resorting to picture or video coding primitives, e.g. JPEG or H.264/AVC, and transmit it to a central server that extracts local features and performs a given visual analysis task. According to $\CTA$, the pixel-level representation of the acquired visual content is actually sent to the sink node. A number of applications rely on compact representations of the content, in the form of local or global features. In this context, $\CTA$ might not be the most efficient approach, since unnecessary and possibly redundant information is sent on the network. Furthermore, the central processing node receives and exploits a lossy version of the originally acquired visual content. Artifacts introduced by coding algorithms may affect the accuracy of several applications~\cite{CTA_ATC}. Several works in the literature aim at adapting both image~\cite{Chao2011} and video~\cite{Agrafiotis2006, Chiang2010} compression architectures so that the quality local features is preserved. 

An alternative paradigm, namely \emph{Analyze-Then-Compress} ($\ATC$), has been introduced in~\cite{CTA_ATC}. Such an approach aims at tackling the limitations posed by $\CTA$. According to $\ATC$, the sensing nodes acquire the visual content, extract information in the form of local or global features, that are encoded and transmitted to a sink node that performs visual analysis based on such features. Such paradigm moves part of the computational complexity from the central unit to the sensing nodes. To this end, efficient algorithms for visual feature extraction~\cite{BAMBOO_ICASSP, Trzcinski13a} and coding architectures tailored to global and local visual features~\cite{Baroffio_VideoSIFT_TIP, Baroffio_VideoBRISK_ICIP, RedondiBACT:ICIP2013, Makar2013} have been proposed. The task efficiency is improved, since only relevant information is actually transmitted over the network. Still, the sink node is not able to reconstruct the original pixel-level representation of the visual content.

In this paper we propose 
a novel hybrid approach to distributed visual analysis tasks aimed at overcoming the limitations of both $\ATC$ and $\CTA$. 
\emph{Hybrid-Analyze-Then-Compress} (HATC) represents an efficient solution for the joint coding of both pixel-level and local feature-level representations. Furthermore, the allocation of the bit budget to either visual content or image feature is thoroughly investigated. 

Moulin et al.~\cite{Moulin2014} addressed the problem of jointly encoding pixel-level content and global image features such as either Bag-of-Words histograms or integral channel features in the context of scene classification or pedestrian detection, respectively. Differently, we focus on the joint encoding of visual content and local image features, typically consisting of sets of salient points, along with their descriptors. 

The rest of this paper is organized as follows: Section~\ref{sec:setup} introduces the problem, defining tools and objectives, Section~\ref{sec:algorithm} describes the proposed paradigm, Section~\ref{sec:experiments} is devoted to experimental evaluation. Finally, Section~\ref{sec:conclusions} draws conclusions and discusses future work.

\section{Problem statement}\label{sec:setup}
Let $\Image$ denote an image that is acquired by a sensing node. Such image is processed in order to extract a set of features $\Descr$. To this end, a detector is applied to the image in order to identify interest points. The number of detected keypoints $M = |\Descr|$ depends on both the image content and on the type and parameters of the adopted detector. Then, a keypoint descriptor is computed starting from the orientation-compensated patch surrounding each interest point. Hence, $d_m \in \Descr$ is a local feature, that consists of two components: i) a 4-dimensional vector $\cv_m = [x_m, y_m, \sigma_m, \theta_m]^T$ , indicating the position $(x_m, y_m)$, the scale $\sigma_m$ of the detected keypoint, and the orientation angle $\theta_m$ of the image patch; ii) a $D$-dimensional vector $\dv_m$, which represents the descriptor associated to the keypoint $\cv_m$. According to \emph{Analzyze-Then-Compress}, the set of features $\Descr$ is encoded 
and transmitted to a sink node for further analysis. On the other hand, \emph{Compress-Then-Analyze} would require the acquired image $\Image$ to be encoded
and transmitted to a central unit where it is analyzed. In details, the sink node receives the bitstream and reconstructs a lossy version of the original image $\ImageDec$. Then, similarly to the case of $\ATC$, a set of local descriptors is extracted and exploited to perform a given visual analysis task. However, the image coding process introduces artifacts that may 
affect the extraction of local features and, as a consequence, the task accuracy.

We propose an alternative approach, namely \emph{Hybrid-Analyze-Then-Compress}, that aims at efficiently coding both pixel-domain and feature-domain representations of the visual content. In particular, according to such paradigm, the decoder is capable of reconstructing both a lossy representation of the original image $\ImageDec$ (encoded with $R_{\ImageDec}$ bits) and a subset of the original features $\Descr_{\HATC}$ (encoded with $R_{\Descr_{\HATC}}$ bits), thus requiring $R_{\HATC} = R_{\ImageDec} + R_{\Descr_{\HATC}}$ in total. 



The $\HATC$ approach is generally applicable to any kind of local feature. In this paper, we focus on the case in which binary descriptors are used, i.e., $\dv_m \in \{0, 1\}^D$. 
Each descriptor element is a bit, representing the result of a pairwise comparison of smoothed pixel intensities sampled from an ad-hoc designed pattern around a given interest point. In particular, we consider BRISK~\cite{DBLP:conf/iccv/LeuteneggerCS11} binary features. 
\section{HATC coding architecture}\label{sec:algorithm}

\begin{figure*}[t]
	\centering
	\subfigure[]{\includegraphics[trim=0cm 0cm 0cm 0cm, clip=true,width=0.45\textwidth]{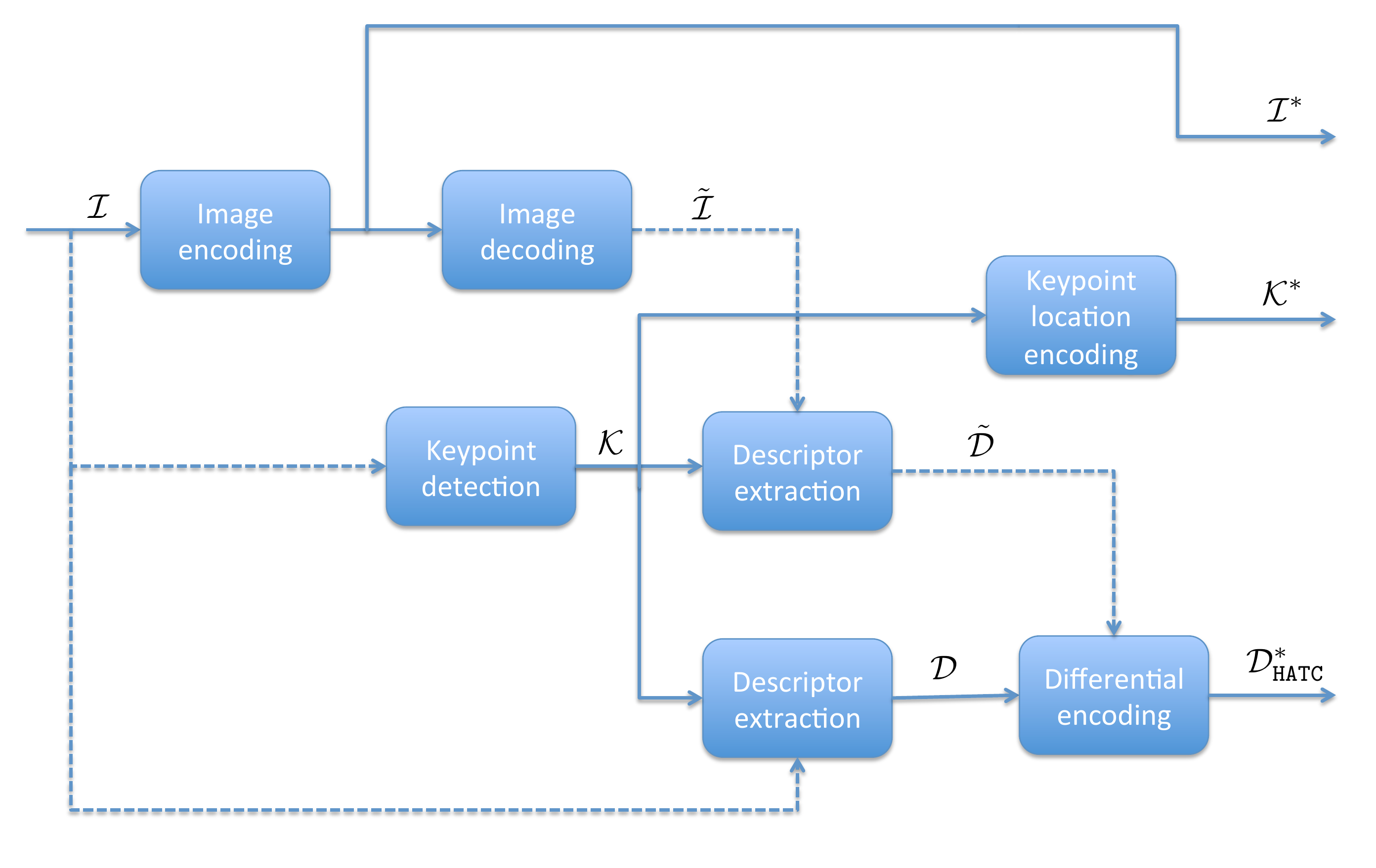}\label{fig:Pipeline_encoder}}
	\subfigure[]{\includegraphics[trim=0cm -2cm 0cm 0cm, clip=true,width=0.26\textwidth]{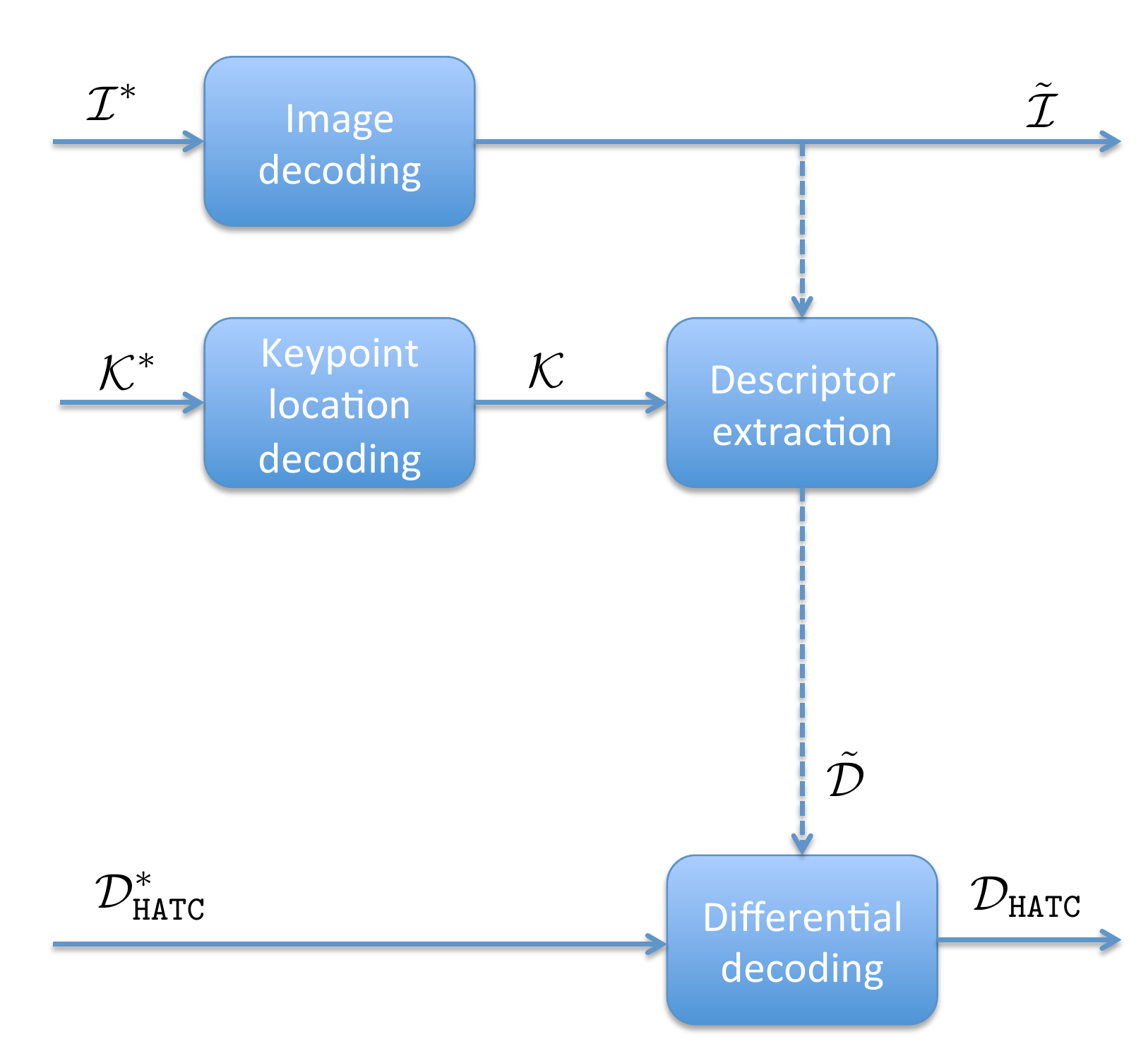}\label{fig:Pipeline_decoder}}
	\caption{Block diagram of a) $\HATC$ joint feature-image encoder; b) $\HATC$ joint feature-image decoder.}
	\label{fig:pipeline} 
\end{figure*}

Figure~\ref{fig:pipeline} illustrates the pipeline of the $\HATC$ coding architecture. As regards the coding of the pixel-level representation of the visual content, $\HATC$ is equivalent to the $\CTA$ approach. That is, the acquired image is encoded and sent to the sink node. Here, the bitstream is decoded and a lossy representation of the image $\ImageDec$ is reconstructed. $\CTA$ would run a detector and a descriptor algorithm on $\ImageDec$, obtaining visual features whose effectiveness is possibly impaired by the image coding artifacts. The key idea behind $\HATC$ is to add an enhancement layer that allows the central processing node to reconstruct a subset $\Descr_{\HATC}$ of the original local descriptors $\Descr$. Such approach allows for the refinement of an arbitrarily-sized subset of features extracted from lossy pixel-level content, yielding a tradeoff between bitrate and task accuracy. The higher the number $Z$ of features that are refined, the higher the resulting bitrate and the higher the accuracy of the visual analysis task to be performed. 

To construct the feature enhancement layer, the sensing node extracts a set of interest points $\Keypoints$ from the acquired image $\Image$. The sensing node computes the sets of descriptors $\Descr$ and $\Descrt$ from the original image $\Image$ and $\ImageDec$, respectively. Descriptors are computed in correspondence to the locations defined by the set $\Keypoints$. Finally, a subset $\Descr_{\HATC}$ of the set of original descriptors $\Descr$ is differentially encoded with respect to the set of lossy descriptors $\Descrt$.

At the central processing node, a lossy representation $\ImageDec$ of the original image is decoded, along with the set of keypoint locations $\Keypoints$. The set of descriptors $\Descrt$ is computed exploiting the lossy coded image $\ImageDec$, at the locations defined by $\Keypoints$. Finally, the bitstream related to the enhancement layer $\Descr^*$ is decoded and exploited in order to reconstruct the subset $\Descr_{\HATC}$ of the original descriptors $\Descr$.

The $\HATC$ paradigm requires three main components to be encoded and transmitted to the central node:

\begin{itemize}
	\item $\Image^*$, i.e., the bitstream needed to reconstruct a lossy representation $\ImageDec$ of the original image $\Image$;
	\item $\mathcal{K}^*$, i.e., the bitstream needed to reconstruct the location of the keypoints extracted from the original image $\Image$;
	\item $\Descr^{*}_{\HATC}$, i.e., the bitstream needed to reconstruct the feature enhancement layer $\Descr_{\HATC}$.
	
\end{itemize}

In a summary, $\HATC$ offers advantages with respect to both $\ATC$ and $\CTA$. First, differently from $\ATC$, the central unit is capable of reconstructing the pixel-level visual content. Second, differently from $\CTA$, $\HATC$ allows the sink node to operate on high quality visual features, yielding a higher task accuracy. 

\subsection{Differential coding of binary local features}

For $\HATC$ to be competitive with other approaches, an effective ad-hoc coding architecture has to be developed. Consider the sets of descriptors $\Descr$ and $\Descrt$, extracted from an input image $\Image$ and its lossy counterpart $\ImageDec$, respectively. The proposed differential coding architecture aims at efficiently encoding the descriptors $\Descr_{\HATC}$, exploiting $\Descrt$ as a predictor. The key tenet behind $\HATC$ is that the two sets of descriptors, extracted in correspondence of a common set of interest point locations, are correlated. In a sense, such a scenario is similar to that of features extracted from contiguous frames of a video sequence. In that case, inter-frame predictive coding can be exploited to improve coding efficiency, reducing the output bitrate~\cite{Baroffio_VideoSIFT_TIP, Baroffio_VideoBRISK_ICIP, BaroffioRCTT:ICIP2013}.

In the case of $\HATC$, given a binary descriptor $\dv_m \in \Descr$ and its counterpart $\dvt_m \in \Descrt$ extracted from the original and the decoded images, respectively, the prediction residual can be computed as 
\begin{equation}
	\cv_m = \dv_m \oplus \dvt_m,
\end{equation} 
that is, the bitwise $XOR$ between $\dv_m$ and $\dvt_m$.

In binary descriptors, each element represents the binary outcome of a pairwise comparison between smoothed pixel intensities. Hence, the dexels (descriptor elements) are potentially statistically dependent, and so are the elements of the prediction residual $\cv_m$. In this context, it is possible to model the prediction residual as a binary source with memory. Let $\pi_j$, $j \in [1,\Dim]$ represent the $j$-th element of a prediction residual, where $\Dim$ is the dimension of such a descriptor. The entropy of such an element can be computed as
\begin{equation}
	H(\pi_{j}) = -p_{j}(0)\log_2(p_{j}(0)) -p_{j}(1)\log_2(p_{j}(1)),
\end{equation}
where $p_j(0)$ and $p_j(1)$ are the probability of $\pi_j = 0$ and $\pi_j = 1$, respectively. Similarly, the conditional entropy of element $\pi_{j_1}$ given element $\pi_{j_2}$ can be computed as
\begin{equation}
	H(\pi_{j_{1}}|\pi_{j_{2}}) = \sum_{x \in \{0,1\}, y \in \{0,1\}} p_{j_{1},j_{2}}(x,y) \log_{2}\frac{p_{j_{2}}(y)}{p_{j_{1},j_{2}}(x,y)},
\end{equation}
with $j_1, j_2 \in [1,\Dim]$. Let $\tilde{\pi}_{j}$, $j = 1,\dots,\Dim$, denote a permutation of the prediction residual elements, indicating the sequential order used to encode a descriptor. The average code length needed to encode a descriptor is lower bounded by
\begin{equation}
	R = \sum_{j=1}^{P}H(\tilde{\pi}_{j}|\tilde{\pi}_{j-1}, \ldots, \tilde{\pi}_{1}).
\end{equation}
In order to maximize the coding efficiency, we aim at finding the permutation of elements $\tilde{\pi}_{1}, \ldots, \tilde{\pi}_{\Dim}$ that minimizes such a lower bound. For the sake of simplicity, we model the source as a first-order Markov source. That is, we impose $H(\tilde{\pi}_{j}|\tilde{\pi}_{j-1}, \dots \tilde{\pi}_{1}) = H(\tilde{\pi}_{j}|\tilde{\pi}_{j-1})$. Then, we adopt the following greedy strategy to reorder the elements of the prediction residual:
\begin{equation}
	\tilde{\pi}_{j} = \begin{cases} 
						\arg \min_{\pi_{j}} H({\pi_{j}}) & j = 1 \\  
						\arg \min_{\pi_{j}} H({\pi_{j}}|\tilde{\pi}_{j-1}) & j \in [2, \Dim]
					  \end{cases}
\end{equation}

Note that such optimal ordering is computed offline, thanks to a training phase, and shared between both the encoder and the decoder.

\subsection{Coding of keypoint locations}

Consider an $N_x \times N_y$ image $\Image$. The coordinates of each keypoint $\cv_m \in \Keypoints$ (at quarter-pel accuracy) are encoded using $\Rate_{\cv_m} = M_n (\log_2 4N_x + \log_2 4N_y + S)$ bits, where $S$ is the number of bits used to encode the scale parameter. Higher coding efficiency is achievable implementing ad-hoc lossless or lossy coding schemes to compress the coordinates of the keypoints~\cite{Tsai:2009:LCM:1653543.1653553}\cite{Tsai:2012:Improved}.

\section{Experiments}\label{sec:experiments}

The effectiveness of the proposed paradigm has been evaluated and compared with that of both \emph{Compress-Then-Analyze} and \emph{Analyze-Then-Compress}, with respect to a content-based image retrieval application. 
\subsection{Datasets}
We exploit the publicly available \emph{Zurich building dataset} (ZuBuD)~\cite{shao03zubud} in order to evaluate the performance of $\HATC$. Such a dataset consists of 1005 pictures representing 201 different Zurich buildings (5 different views for each object). A test set composed of 115 image queries, each one capturing a different building, is also provided. Database and query images have heterogeneous resolutions and imaging conditions.
As regards the training phase, 1000 images have been randomly sampled from the MIRFLICKR~\cite{huiskes08} dataset and they have been exploited to compute the coding-wise optimal dexel order and the associated coding probabilities, as illustrated in Section~\ref{sec:algorithm}.
\subsection{Methods}
We compared the performance of the following paradigms:
\begin{itemize}
	\item \emph{Compress-Then-Analyze} ($\CTA$): each query picture is encoded resorting to JPEG. Subsequently, BRISK local features are extracted from the lossy compressed image and exploited for the retrieval pipeline;
	\item \emph{Analyze-Then-Compress} ($\ATC$): each query picture is processed in order to extract a set of BRISK features, that are encoded resorting to the architecture proposed in~\cite{RedondiBACT:ICIP2013} and exploited for the retrieval pipeline;
	\item \emph{Hybrid-Analyze-Then-Compress} ($\HATC$): a local feature enhancement layer, composed by a subset of the BRISK feature extracted from the uncompressed image, is generated and differentially encoded according to the procedure presented in Section~\ref{sec:algorithm}. Such features are exploited for the retrieval pipeline.
	\end{itemize}
\subsection{Parameter settings}
As for $\CTA$, we define a set of possible values for the JPEG quality factor $Q = \{5, 10, 15, 20, 50, 70\}$ in order to generate a rate-accuracy curve. As to $\ATC$, a similar rate-accuracy curve is obtained by imposing different BRISK detection thresholds $t_{BRISK} = \{70, 75, 80, 85, 90, 95, 100, 105\}$. Finally, as to $\HATC$, for each JPEG quality factor, a rate-accuracy curve is obtained by setting the number $Z = \{25, 50, 100, 150\}$ of features to be refined resorting to a feature enhancement layer, as reported in Section~\ref{sec:algorithm}.
\subsection{Evaluation metrics}
We evaluate the performance in terms of rate-accuracy curves. In particular, the accuracy of the task is evaluated according to the \emph{Mean Average Precision} (MAP) measure. Given an input query image $\Image_q$, it is possible to define the \emph{Average Precision} as 
\begin{equation}
	AP_{q} = \frac{\sum_{k=1}^Z P_{q}(k)r_{q}(k)}{R_{q}},
\end{equation}
where $P_{q}(k)$ is the precision (i.e., the fraction of relevant documents retrieved) considering the top-$k$ results in the ranked list of database images; $r_{q}(k)$ is an indicator function, which is equal to 1 if the item at rank $k$ is relevant for the query, and zero otherwise; $R_{q}$ is the total number of relevant document for query $\Image_q$ and $Z$ is the total number of documents in the list.
The overall \emph{Mean Average Precision} for the whole set of query images is computed as
\begin{equation}
	MAP = \frac{\sum_{q = 1}^Q AP_{q}}{Q}, 
\end{equation}
where $Q$ is the total number of queries.

The quality of a JPEG coded image is evaluated according to its PSNR with respect to the uncompressed image.

\subsection{Results}

\begin{figure}[t]
	\centering
	\includegraphics[trim=0cm 0cm 0cm 0cm, clip=true,width=0.45\textwidth]{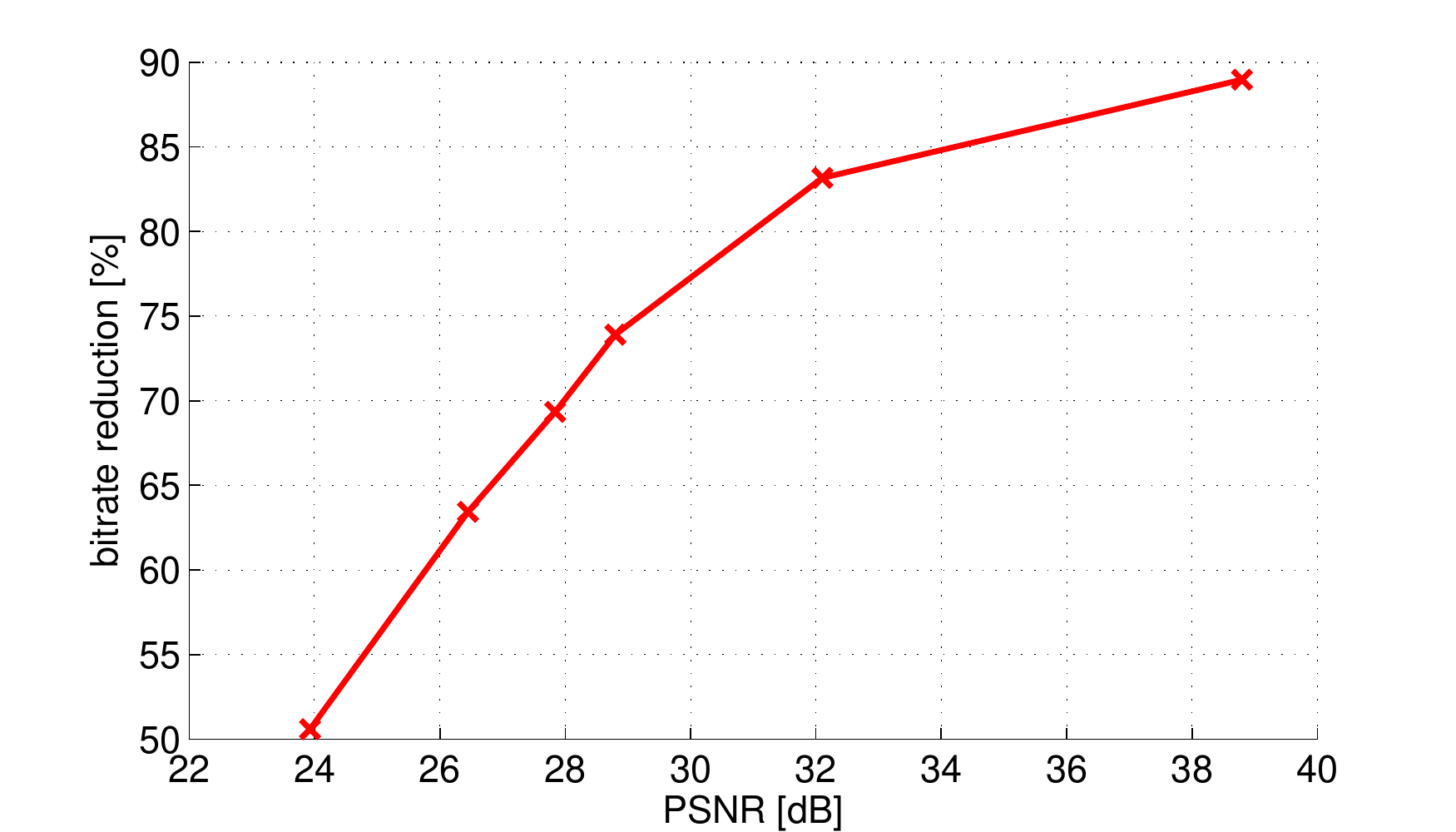}\label{fig:homog}
	\caption{Feature coding efficiency as a function of the distortion (PSNR) between the original and the lossy pixel-level visual content.}
	\label{fig:HATC_gain} 
\end{figure}

\begin{figure}[t]
	\centering
	\includegraphics[trim=0cm 0cm 0cm 0cm, clip=true,width=0.45\textwidth]{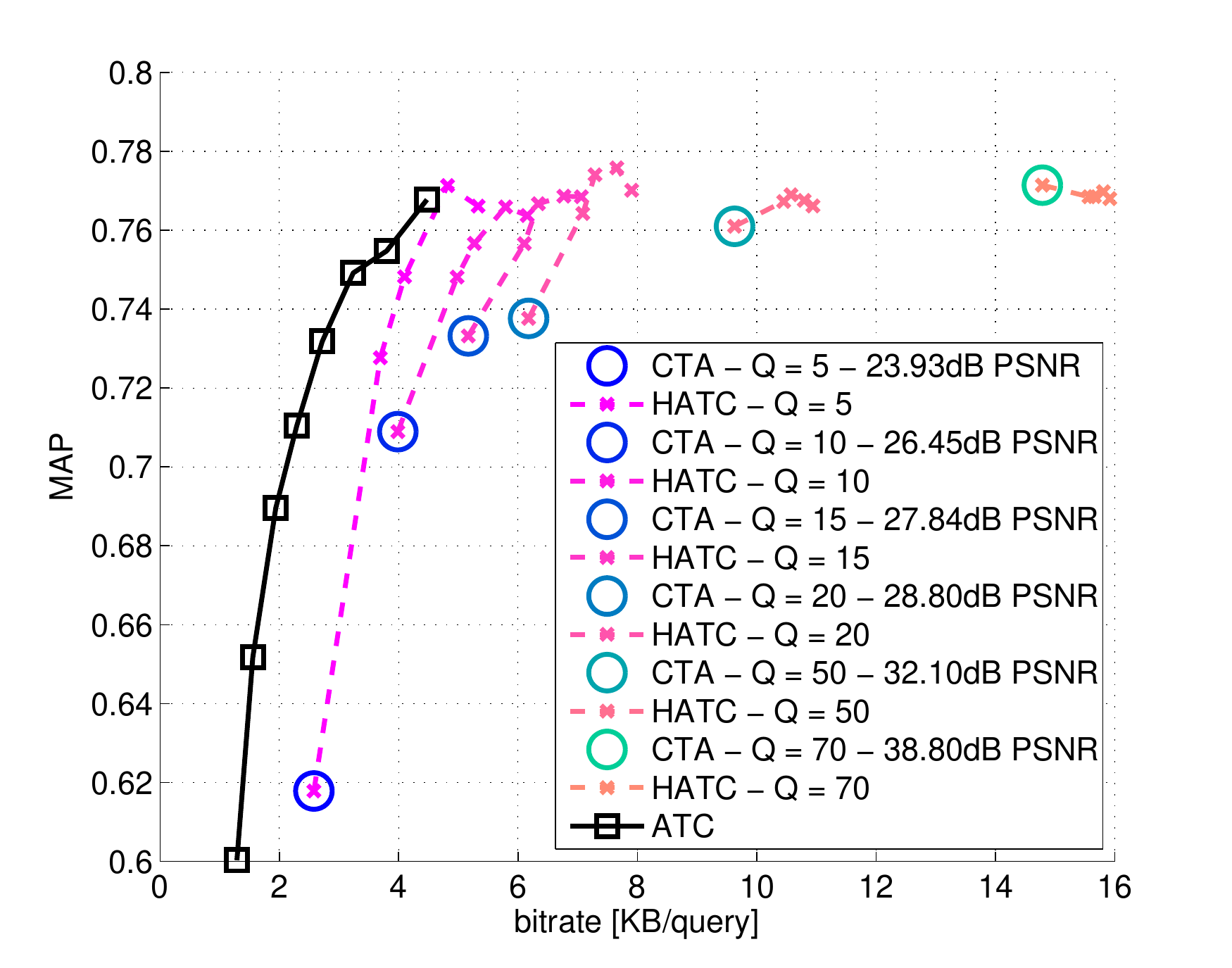}\label{fig:homog}
	\caption{Rate-accuracy curves comparing the performance of $\ATC$, $\CTA$ and $\HATC$.}
	\label{fig:all_approaches} 
\end{figure}

\begin{figure}[t]
	\centering
	\includegraphics[trim=0cm 0cm 0cm 0cm, clip=true,width=0.45\textwidth]{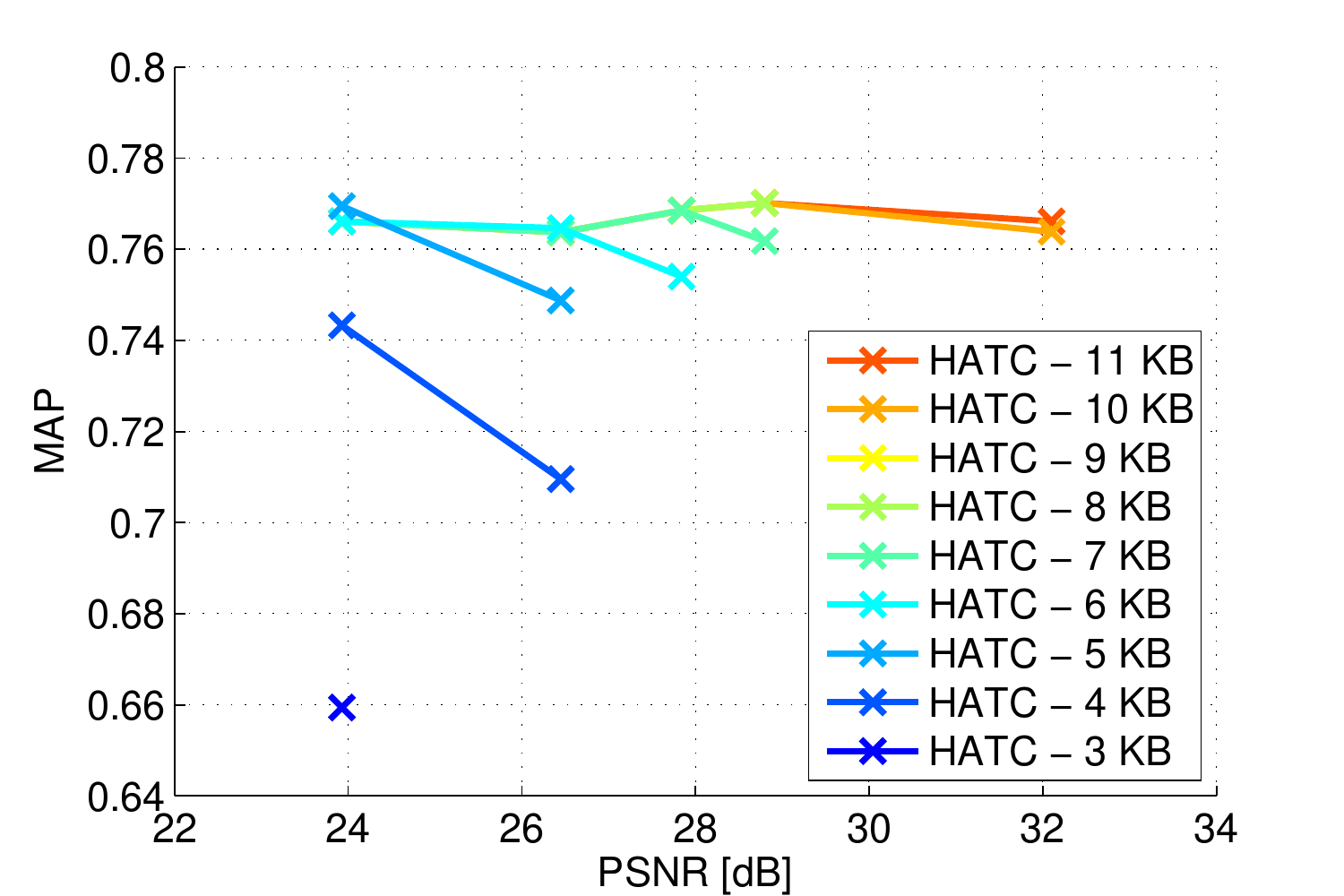}\label{fig:homog}
	\caption{Tradeoff between pixel-level distortion (PSNR) and visual analysis task accuracy (MAP) obtained resorting to the $\HATC$ architecture. Each curve refers to a target bitrate budget.}
	\label{fig:MAP_PSNR} 
\end{figure}

Figure~\ref{fig:HATC_gain} shows the feature coding efficiency achieved by the differential encoding module (see Figure~\ref{fig:pipeline}) as a function of the distortion (PSNR) between the original image and the lossy one reconstructed resorting to $\CTA$. The lower the distortion (the higher the PSNR), the more effective the $\HATC$ feature coding architecture. Nonetheless, high PSNRs correspond to low distortion values, and thus the accuracy increment yield by $\HATC$ is smaller. 

Figure~\ref{fig:all_approaches} compares the rate-accuracy performance of the three approaches. For example, when 4 KB/query are allocated, $\CTA$ achieves a MAP equal to 0.71. This value increases to 0.75 when using $\HATC$, trading-off accuracy for visual quality (which decreases from 26.4dB to 23.9dB). $\ATC$ achieves a slightly higher MAP (0.76), but the pixel-domain content is not available at the decoder. A similar analysis can be performed for different target bitrate budgets. Figure~\ref{fig:MAP_PSNR} shows the MAP-PSNR trade-offs that are achievable when targeting a given bitrate. When the available bitrate is equal to 3KB per query, a single working point corresponding to 0.66 MAP @ 24dB PSNR is achievable. At higher target bitrates (e.g. 4-7 KB/query), it is possible to select a trade-off between MAP and PSNR by accurately allocating the available bitrate to either the pixel-level or the feature-level representations. 

\section{Conclusions}\label{sec:conclusions}
In this paper we propose \emph{Hybrid-Analyze-Then-Compress}, an effective paradigm tailored to distributed visual analysis tasks. Such model exploits a joint pixel- and local feature-level coding architecture, leading to significant bitrate savings. Future work will aim at improving the coding efficiency of both the keypoint location and the descriptor enhancement layer modules and at extending the approach to different classes of local features (e.g. SIFT, SURF descriptors).

\balance

\bibliographystyle{IEEEbib}
\bibliography{IEEEabrv,string-defs,bibfile}

\end{document}